\documentclass[aps,prb,reprint,showpacs,groupedaddress]{revtex4-1}

\usepackage{bm}
\usepackage{wasysym}
\usepackage{amssymb}
\usepackage[cp1251]{inputenc}
\usepackage[dvips]{graphicx}
\usepackage[usenames]{color}
\usepackage{hyperref}
\begin{document}

\title{Electrically controllable cyclotron resonance}

\author{A. A. Zabolotnykh}
\affiliation{Kotelnikov Institute of Radio-engineering and Electronics of the RAS, Mokhovaya 11-7, Moscow 125009, Russia}

\author{V. A. Volkov}
\email{Volkov.V.A@gmail.com}
\affiliation{Kotelnikov Institute of Radio-engineering and Electronics of the RAS, Mokhovaya 11-7, Moscow 125009, Russia}

\date{\today}

\begin{abstract}
Cyclotron resonance (CR) is considered one of the fundamental phenomena in conducting systems.
In this paper, we study CR in a gated two-dimensional (2D) electron system (ES). Namely, we analyze the absorption of electromagnetic radiation incident normal to the gated 2DES, where a standard dielectric substrate separates the 2D electron sheet and the metallic steering electrode (''gate''); the whole system is placed in the perpendicular magnetic field.
Our analysis reveals the redshift of the absorption peak frequency compared to the electron cyclotron frequency. The redshift appears in low-frequency regime, when the resonant frequency is much less than the frequency of Fabry-Perot modes in natural resonator ''2D electron sheet -- substrate -- gate”. 	
Moreover, we find this shift to be dependent on the electron density of 2DES. Therefore, it can be controlled by varying the gate voltage. We predict that the shift can be large in realistic gated or back-gated 2DESs. The obtained controllability of CR in gated 2DES opens the door for exploring new physics and applications of this phenomenon.
\end{abstract}

\maketitle
\section{Introduction}
In a classical plasma, an electron exposed to the magnetic field $\bm B$ rotates in a closed orbit with the cyclotron frequency $\omega_c =eB/mc $, where $-e$ and $m$ are the electron charge and effective mass, and $c$ is the speed of light in vacuum (we note that in this paper, we use the Gauss units). The absorption of an electromagnetic wave by an electron system placed in the magnetic field, known as cyclotron resonance (CR), has long been used in fundamental studies and applications of both non-degenerate gaseous and degenerate condensed-matter plasmas. In two-dimensional electron systems (2DESs), CR was observed initially in silicon inversion layers \cite{Abstreiter1974, Allen1974}. Nowadays, it is used extensively to characterize various kinds of 2DESs.

Depending on the presence of a steering electrode (the gate), all 2DESs can be divided into gated and ungated systems. Theoretically, the CR in infinite ungated 2DES was studied in Ref.~\cite{Chiu1976}, see also \cite{Mikhailov2004,Rodionov2019}. It was found that for the 2DES situated between two half-spaces with dielectric permittivities $\varkappa_1$ and $\varkappa_2$, the resonance frequency equals $\omega_c$. At the same time, the half-width of the CR line is defined by the sum of collisional broadening $\gamma=1/\tau$, with $\tau$ being the electron relaxation time in 2DES, and collective radiative broadening $\Gamma_{\varkappa}=2 \Gamma /(\sqrt{\varkappa_1}+\sqrt{\varkappa_2})$, where
\begin{equation}
  \Gamma =\frac{2\pi e^2 n}{mc},
\end{equation}
and $n$ is the 2D electron concentration.
In ungated GaAs quantum wells, the collective nature of the CR decay rate $\Gamma_{\varkappa}$ has been observed in time-domain experiments \cite{Zhang2014, Manfra2016}, and interpreted as the effect of superradiance \cite{Dicke1954, Mazza2019}. 

However, to the best of our knowledge, the absorption of electromagnetic radiation by a gated 2DES placed in the perpendicular magnetic field, i.e., CR in gated 2DESs, has not been previously studied in detail. 
The gated 2DES has a unique feature --- a technical possibility to control the electron density over a wide range through the gate voltage. Thus, it enables, for example, fine-tuning of the electron-electron interaction parameter. 

\begin{figure}[t]
	\includegraphics[width=7.0cm]{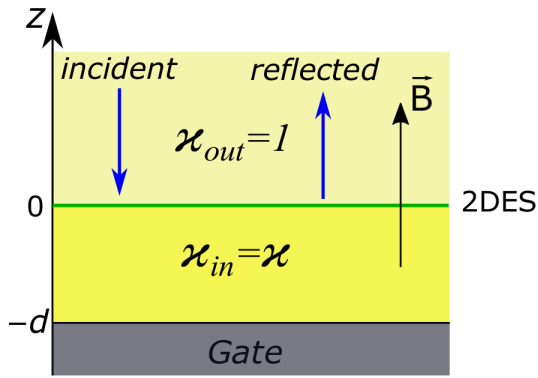}
	\caption{ \label{Fig:geom_abs} Diagram of the system under consideration: 
	2D electron sheet ($z=0$), a dielectric substrate ($-d < z < 0$) with permittivity $\varkappa$, and the metallic back-gate ($z \le-d$) form an analog of the Fabry-Perot resonator.
	}
\end{figure}

It should also be noted that gated 2DES is one of the simplest structures for studying the light-matter coupling since the metal gate, the dielectric substrate, and the 2DES form a Fabry-Perot resonator (see Fig.~\ref{Fig:geom_abs}) for the photon modes which interact with the electrons in the 2DES.
Over the past decade, the interaction between the cyclotron motion of electrons and discrete photon modes in specially designed microwave or THz resonators has been attracting great interest having been investigated extensively in various independent studies \cite{Muravev2011,Muravev2013,Scalari2012,Maissen2014,Paravicini2019, Bayer2017, Li2018, Abdurakhimov2016,Geiser2012}, for review see \cite{FornDiaz2019}. It has been found that the light-matter coupling leads to the modification of the response, though, typically, within a finite range of magnetic fields corresponding to the anti-crossing of the cyclotron frequency and the photon frequency of the resonator. 

In the present paper, we report on significant and unexpected CR ''renormalization'' in a gated 2DES --- namely, the shift of the resonant frequency away from the cyclotron frequency $\omega_c$ and the narrowing of the resonance linewidth. We emphasize that, surprisingly, it takes place in the low-frequency regime, when the cyclotron frequency is much less than the characteristic frequency of the resonator formed by the 2DES and the metal gate, with a dielectric substrate in between. Furthermore, we assert that the resultant frequency shift can be large, even in conventional 2D electron structures with the back-gate. Therefore, it can be rather easily observed experimentally.

We establish the CR in a gated 2DES to be governed by the retardation parameter $A$ defined as the ratio of the characteristic velocity in a gated 2DES ($V_p$) to the speed of light in the dielectric substrate separating the 2DES and the metal gate ($c/\sqrt{\varkappa}$). Formally, the characteristic velocity equals to that of plasma waves in the gated 2DES \cite{Chaplik1972}:
\begin{equation}
	V_p = \sqrt{\frac{4 \pi n e^2 d}{\varkappa m}},
\end{equation}
where $d$ is the distance between the gate and 2DES. The retardation parameter can be written in three equivalent forms, as follows:
\begin{equation}
	\label{retard}
	A^2= V_p^2\varkappa/c^2 = 2d\Gamma/c = 4 \pi  n e^2 d /(mc^2).
\end{equation}
We find that the resonant frequency can be expressed in terms of $A$ through a simple relation:
\begin{equation}
\label{res}
	\omega_{res} = \omega_c /(1+A^2).
\end{equation}
It should be stressed that the given retardation parameter (\ref{retard}) can be easily modified by changing electron concentration by the chemical doping or with the gate voltage, which enables the electrical control of the CR in gated 2DESs.

\section{Approach and key equations}
To determine the CR frequency and linewidth, we consider the absorption of the electromagnetic plane wave incident normally onto the gated 2DES, as depicted in Fig.~\ref{Fig:geom_abs}. Let the $\delta$-thin 2DES and the top surface of an ideal metal gate be situated, respectively, at $z=0$ and $z=-d$, with dielectric permittivities of the substrate ($-d < z < 0$) and the medium above the 2DES ($z > 0$) equal $\varkappa$ and unity, accordingly. Also, consider the system in a constant magnetic field $\bm{B}$ directed along the $z$-axis. 

To calculate the absorption, we follow the classic approach based on Maxwell's equations and Ohm's law $\bm{j}=\widehat{\sigma}(\omega) \bm{E}$, with $\bm{j}$ and $\widehat{\sigma}(\omega)$ denoting, correspondingly, the current density and the dynamical conductivity tensor of the 2DES; and $\bm{E}$ being the electric field in the 2DES plane. It should be mentioned that this is a standard approach that is widely used to determine the response of different 2D structures, see, for example, Refs.~\cite{Chiu1976,Mikhailov2005,Mikhailov2006,Kukushkin2003,Kovalskii2006, Satou2007, Gusikhin2020,Oriekhov2020,Nikulin2021}.

At $z>0$ there exist the incident and reflected electromagnetic waves, which have the forms of $\bm{E}_i\exp(-i\omega z/c-i\omega t)$ and $\bm{E}_r\exp(i\omega z/c-i\omega t)$. At the same time, within the substrate, for $-d<z<0$, there are waves propagating likewise in the positive and negative directions along the $z$-axis. According to standard boundary conditions for $\bm{E}(z)$ (which, in our case, lies in the $z=const$ plane), the electric field vanishes at the surface of the metal gate $z=-d$, and is continuous at the 2DES plane $z=0$, whereas the discontinuity in the $z$-derivative of $\bm{E}(z)$ can be defined as follows:
\begin{equation}
	\partial_z \bm{E}(z)|^{z=+0}_{z=-0}=-\frac{4\pi i\omega}{c^2} \bm{j}.
\end{equation}
Here, the last condition arises from equations $\partial_z E_y=-i\omega H_x/c$ and $\partial_z E_x=i\omega H_y/c$, as well as the discontinuity of the magnetic field components $H_x$ and $H_y$ due to the presence of 2D current $\bm j$: $H_x(z)|^{+0}_{-0}=4\pi j_y/c$ and $H_y(z)|^{+0}_{-0}=-4\pi j_x/c$.

As a matter of convenience, we introduce the ''circular'' variables: $E_{i\pm}=E_{ix}\pm i E_{iy}$, $E_{r\pm}=E_{rx}\pm i E_{ry}$, and $\sigma_{\pm}=\sigma_{xx}\mp i\sigma_{xy}$, where $\sigma_{xx}$ and $\sigma_{xy}$ are, respectively, the dynamical longitudinal and transverse (Hall) conductivities of the 2DES. Then, the 2DES response can be found separately for each circular polarization. Using the boundary conditions mentioned above, the amplitude of the reflection coefficient $r_{\pm}=E_{r\pm}/E_{i\pm}$ can be expressed as:
\begin{equation}
\label{reflection}
 r_{\pm}(\omega)=\frac{1- i\sqrt{\varkappa}\cot(\omega\sqrt{\varkappa}d/c)-4\pi\sigma_{\pm}/c} {1+ i\sqrt{\varkappa}\cot(\omega\sqrt{\varkappa}d/c)+4\pi\sigma_{\pm}/c},
\end{equation}
where ''$+$'' and ''$-$'' signs correspond to different circular polarizations of the incident wave.

Now, let us find the shape of the resonance line from the calculation of the energy absorption coefficient $P_{\pm}(\omega)=1-|r_{\pm}|^2$. 
However, to obtain the explicit expression for $P_{\pm}(\omega)$, one needs to specify the model for the 2DES conductivity tensor $\widehat{\sigma}(\omega)$. In our further analysis we use a simple Drude model. Then, in the framework of the model, $\sigma_{xx}$ and $\sigma_{xy}$ can be expressed as follows:
\begin{eqnarray}
	\label{Drude}	
	\sigma_{xx}	=\frac{e^2n}{m}\,\frac{\gamma-i\omega}{(\gamma-i\omega)^2+\omega_c^2},\\ \sigma_{xy}=\frac{e^2n}{m}\,\frac{-\omega_c}{(\gamma-i\omega)^2+\omega_c^2},\nonumber
\end{eqnarray}
where $\gamma=1/\tau$ is the inverse electron relaxation time, which is assumed to be constant.

Substituting Eqs.~(\ref{Drude}) into Eq.~(\ref{reflection}), we find the exact expression for the energy absorption coefficient $P_{\pm}(\omega)$, given the circular polarization of the incident wave:
\begin{widetext}
\begin{equation}
	\label{Abs}
	P_{\pm}(\omega)=\frac{8\Gamma \gamma}{\left(\sqrt{\varkappa} (\omega\pm\omega_c)\cot\left(\frac{\omega d\sqrt{\varkappa}}{c}\right)+2\Gamma\right)^2+\gamma^2\left(\varkappa\cot^2\left(\frac{\omega d\sqrt{\varkappa}}{c}\right)+1\right)+ 4\Gamma\gamma	+(\omega\pm\omega_c)^2},
\end{equation}
\end{widetext}

Importantly, of particular interest to us is the low-frequency regime $\omega\ll c/(d\sqrt{\varkappa})$, where we can neglect all but the dominant terms in the full expression for $P_{\pm}(\omega)$ (\ref{Abs}) to obtain the simplified relation: 
\begin{eqnarray}
\label{Abs_as}
	P_{\pm}(\omega)=\frac{4\gamma A^2d/c}{(\frac{\omega\pm\omega_c}{\omega}+A^2)^2 +\frac{\gamma^2}{\omega^2}+
		\frac{2\gamma  A^2 d}{c}+\frac{ d^2 (\omega\pm\omega_c)^2}{c^2}},
\end{eqnarray}
which can be used to find the resonant frequency $\omega_{res}$ and linewidth $\Delta\omega$ analytically. 

\section{Radiation absorption analysis}
To be specific, we consider the case of $\omega>0$ and $\omega_c>0$ to analyze $P_-(\omega)$, which corresponds to the active circular polarization of the incident wave. It should be noted that for the passive polarization, we do not find any resonance response.

From Eq.~(\ref{Abs_as}) we obtain that in classically strong magnetic field ($\gamma\ll \omega_c$), the resonant frequency takes the form described in (\ref{res}), i.\,e., the resonance frequency is shifted from $\omega_c$ by the parameter $(1+A^2)^{-1}$.

Also, provided that $\gamma\ll \omega_c$, the linewidth of the resonance becomes:
\begin{equation}
\label{LW}
	\Delta\omega=2\frac{\gamma+A^2 \omega_{res}^2d/c}{1+A^2}.
\end{equation}
Here, the total linewidth $\Delta\omega$ is the sum of ''renormalized'' collisional broadening $2\gamma/(1+A^2)$ and radiative broadening $2A^2 \omega_{res}^2d/c/(1+A^2)$.

\begin{figure}
			\includegraphics[width=7.5cm]{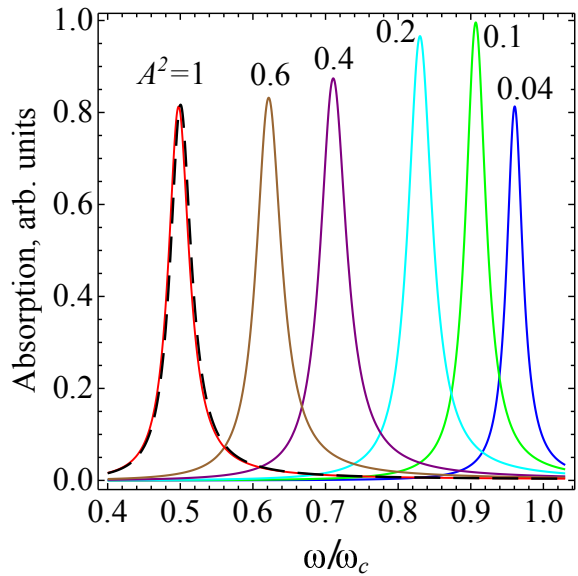}
			\caption{ \label{Fig:Absorption_w} Absorption in the gated 2DES, calculated from the exact relation (\ref{Abs}) as a function of the radiation frequency $\omega$, given $\gamma/\omega_c=0.01$, $\varkappa=12.8$, and $c/(d\omega_c)=10$. Solid lines correspond to indicated values of $A^2=0.04,0.1,0.2,0.4,0.6,1$. Dashed line designates the asymptotic relation of the absorption in (\ref{Abs_as}) for $A=1$ calculated in the limit $\omega d\sqrt{\varkappa}/c\ll 1$.
  }
\end{figure}

\begin{figure}
	\includegraphics[width=7.5cm]{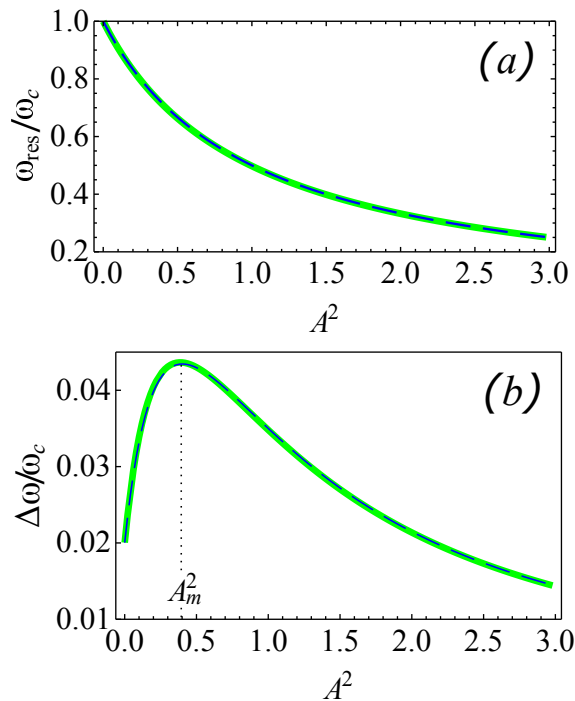}
	\caption{ \label{Fig:Position} Dependence of the position $\omega_{res}$ (a) and linewidth $\Delta\omega$ (b) of the absorption maximum on the parameter $A^2$ (\ref{retard}), computed for $\gamma/\omega_c=0.01$, $\varkappa=12.8$, and $c/(\omega_cd)=10$. Green (solid) lines indicate the exact numerical calculation of $\omega_{res}$ and $\Delta\omega$ from Eq.~(\ref{Abs}); blue (dashed) lines correspond to the analytical expressions (\ref{res}) and (\ref{LW}) for $\omega_{res}$ and $\Delta\omega$, accordingly. The maximum linewidth is reached at the value of $A_m^2$ defined by Eq.~(\ref{LW_max}).
	}
\end{figure}

In Fig.~\ref{Fig:Absorption_w}, we plot the frequency dependence of the absorption in the gated 2DES, according to the exact formulation (\ref{Abs}), where solid lines denote the set of curves for different values of $A^2$. For comparison, the dashed curve indicates the line shape obtained from the asymptotic relation in~(\ref{Abs_as}) for $A=1$.  Clearly, the asymptotic result is in excellent agreement with the exact calculation. In addition, Fig.~\ref{Fig:Position} shows the resonance position and linewidth as a function of the retardation parameter.  Comparing the given numerical and analytical data likewise makes it evident that calculations based on the exact expression for $P_{-}$ (\ref{Abs}) (green solid lines) perfectly agree with the asymptotic results from (\ref{res}) and (\ref{LW}) (blue dashed lines). We also note that when expressed in terms of $A^2$, the linewidth $\Delta\omega$ reaches its maximum value at $A_m^2$ defined by:
\begin{equation}
\label{LW_max}
	A_m^2= \sqrt{\frac{3\omega_c^2 d}{\gamma c} +\frac{\omega_c^4 d^2}{\gamma^2 c^2}}-1-\frac{\omega_c^2 d}{\gamma c}.
\end{equation}

Next, let us analyze the absorption at the resonance frequency, $P_{-}(\omega_{res})$, as a function of the retardation parameter, $A\propto \sqrt{n}$. Considering the extreme cases, we recognize that in the limit of $A\to 0$, the absorption approaches zero, as 2DES becomes virtually absent. Similarly, in the limit of $A\to \infty$, 2DES attains infinite conductivity, which also leads to zero absorption due to full reflection of the incident radiation. Therefore, at finite values of $A$, the absorption is expected to have one or more maxima.
Furthermore, we find that under the condition $\omega_c^2 d/(\gamma c)<4$, i.e. ''weak'' magnetic field, the absorption $P_{-}(\omega_{res})$ exhibits a single maximum at $A = 1$. On the other hand, given $\omega_c^2 d/(\gamma c)>4$, i.e. ''strong'' magnetic field, there occur two local maxima of $P_{-}(\omega_{res})$ at $A^2_{1,2}$ defined as follows:
\begin{equation}
	A^2_{1,2}=-1+\frac{\omega_c^2 d}{2\gamma c}\pm \sqrt{\frac{{\omega_c^4 d^2}}{4\gamma^2 c^2}-\frac{\omega_c^2 d}{\gamma c}},
\end{equation}
with the local minimum of $P_{-}(\omega_{res})$ situated between the maxima, at $A=1$.

\section{Eigen modes analysis}
As for the qualitative interpretation of the obtained results, we consider the electrodynamic eigen modes of the system under study. The complex-valued frequencies of these modes are defined by the poles of the reflection (as well as absorption) coefficient (\ref{reflection}), where the real and imaginary parts of the frequency correspond to oscillations and the decay rate of the mode. To be specific, we give further consideration to the poles of $r_{-}(\omega)$. Hence, given the conductivity tensor~(\ref{Drude}) based on the Drude model, and the Fabry-Perot frequency $\omega_{FP} =c/(\sqrt {\varkappa} d)$, the denominator of $r_{-}(\omega)$ becomes zero provided that
\begin{equation}
\label{poles}
	(\omega -\omega_c +i\gamma)(\cot(\omega/\omega_{FP}) -i/\sqrt{\varkappa})= -A^2\omega_{FP}.
\end{equation}

Let us discuss the structure of the obtained equation. The first bracketed factor on the left-hand side of Eq.~(\ref{poles}) goes to zero at $\omega=\omega_c-i\gamma$, which can be interpreted as single-particle cyclotron motion. The other bracketed factor vanishes at the frequencies of the Fabry-Pero resonator. Hence, the retardation parameter $A$ (\ref{retard}) on the right-hand side of Eq.~(\ref{poles}) relates to the interplay of the two families of modes mentioned above. Increasing $A$ results in the interaction of the cyclotron motion with the photonic modes of the resonator, leading to the ''shift'' of the frequency from $\omega_c-i\gamma$.

Considering a high-quality resonance with $Im\,\omega\ll Re\,\omega$ and $\gamma \ll \omega_c$ in the low-frequency limit $|\omega/\omega_{FP}| \ll 1$, from Eq.~(\ref{poles}) we obtain:
\begin{equation}
\label{complex_fr}
	\omega = \frac{\omega_c}{1+ A^2} -i\frac{\gamma}{1+ A^2} -i \frac{\omega_c^2 A^2 d/c}{(1+ A^2)^3}.
\end{equation}
The real and imaginary parts of the resultant frequency describe, respectively, the position and broadening of the resonance line.
As expected, the expressions for the real and imaginary parts defined by Eq.~(\ref{complex_fr}) are identical, accordingly, to the resonance frequency~(\ref{res}) and the half-width $\Delta\omega/2$~(\ref{LW}) obtained from the absorption analysis.

\section{Discussion and conclusions}
Now, let us compare the absorption of a circular polarized electromagnetic wave in gated and ungated 2DESs exposed to the magnetic field. In the case of the ungated 2DES in vacuum, the absorption maximum appears exactly at the cyclotron frequency $\omega_c$, while the half-linewidth of the resonance equals $\gamma+\Gamma$ ~\cite{Chiu1976}. By contrast, in the gated 2DES, the factor $(1+A^2)^{-1}$ leads to the shift of the absorption peak away from $\omega_c$, as defined in Eq.~(\ref{res}), as well as the reduction in the linewidth. In addition, the radiative contribution to the linewidth becomes significantly suppressed due to the factor $d^2\omega_{res}^2/c^2 \ll 1$, according to Eq.~(\ref{LW}). Thus, in the gated 2DES, one can obtain that when the radiative contribution to the linewidth dominates that of collisional origin, the linewidth narrows with increasing electron concentration: $\Delta\omega\propto A^{-4}\propto n^{-2}$, which is very much unlike the case of the ungated 2DES.

For practical purposes, we estimate the retardation parameter $A$ in 2DES based on a back-gated GaAs/AlGaAs quantum well, given the following characteristic parameters: $d=0.4$~mm, $n= 5\cdot 10^{11}$ cm$^{-2}$, and $m=0.066m_0$, where $m_0$ is the free-electron mass. As a result, we find $A \approx 1$, which in clean samples corresponds to the resonant frequency equal half the cyclotron frequency $\omega_c$. Therefore, the shift in the resonance frequency of standard back-gated semiconductor structures can be far from negligible. However, it should be noted that since our analysis assumes an infinite 2DES, the lateral size of the actual sample must be large relative to the distance $d$ to make our calculations practically applicable.

Let us also consider a possible disorder in the boundary between the metal gate and the dielectric substrate. We believe it can be simulated by the variation in the dielectric thickness $d$, which implies $\delta A$ fluctuation since the retardation parameter $A$ is proportional to the square root of $d$ (\ref{retard}). We maintain that our results, on the whole, are reasonable, provided that $\delta A \ll A$. Likely, the resonance linewidth can have an increase by the parameter $\delta A/ A$. Concerning standard semiconductor structures, for example, those based on GaAs/AlGaAs quantum wells with metal gates, the surface between the dielectric substrate and the gate can be made sufficiently clean and smooth. Typically, in such structures, $d$ varies from 100 nm to 500 $\mu m$, whereas the fluctuation of $d$ does not exceed 1 nm. Therefore, we believe that in given experimental structures, the condition $\delta A\ll A$ is satisfied, and the disorder in the boundary has no significant effect.

Let us analyze the obtained redshift in the resonance frequency in relation to the Kohn's theorem~\cite{Kohn1961}, which can be interpreted as follows: electron systems in the magnetic field cannot have a resonant response at frequencies below the electron cyclotron frequency. It should be mentioned, however, that this theorem can be inapplicable under certain conditions --- for example, due to non-parabolic electron dispersion \cite{Keller2020}, polarons \cite{Hu1996}, ultrasound~\cite{Kukushkin2002}, non-equilibrium and dynamic effects \cite{Maag2016, Mittendorff2015}, etc. In our case of the gated 2DES, the electromagnetic retardation effects (that were not taken into account by the theorem) lead to the redshift of the resonant frequency away from $\omega_c$. It follows from Eqs.~(\ref{retard}) and (\ref{res}) that the frequency shift appears due to $A^2=4 \pi  n e^2 d /(mc^2)$. In the formal non-retarded limiting case of $c \to \infty$, and, therefore, $A\to 0$, the resonance occurs at the cyclotron frequency $\omega_c$, in accordance with Kohn's theorem.

In summary, we have conducted the analytical and numerical investigation of the absorption of electromagnetic wave incident normally onto the gated or back-gated 2DES in the presence of a perpendicular magnetic field. Importantly, the study takes into account the effect of electromagnetic confinement in the natural resonator formed by the 2DES and the metallic back-gate, with a dielectric substrate in between. Unexpectedly, we find the redshift in the resonance frequency (\ref{res}) from the cyclotron frequency $\omega_c$ and narrowing of the linewidth~(\ref{LW}), to occur in the low-frequency regime when the radiation frequency is much smaller than the Fabry-Perot frequency of the resonator. We establish that given effect is controlled by the retardation parameter (\ref{retard}), which depends on the electron concentration in the 2DES and, therefore, can be easily controlled by the gate voltage. We prove that the retardation parameter can be large enough, even in standard back-gated samples. As a result, it can lead to a tremendous shift and narrowing of the CR line in a gated 2DES. Therefore, gated and especially back-gated 2DESs prove very promising for exploring new physical effects, for instance, the experimental studies of the extreme regimes of light-matter coupling.

\begin{acknowledgments}
We would like to thank I. V. Kukushkin and V. M. Muravev for numerous stimulating discussions. 
The work of A.A.Z. was supported by the Foundation for the Advancement of Theoretical Physics and Mathematics ''BASIS'' (project No. 19-1-4-41-1). The work was done within the framework of the state task and supported by the Russian Foundation for Basic Research (project No. 20-02-00817).
\end{acknowledgments}

\end{document}